\documentclass[12pt]{book}

\usepackage[dvips]{graphicx,color}
\usepackage{makeidx,tsukuba}
\makeauthorindex
\makeindex

\begin{document}

\BookTitle{\itshape The 28th International Cosmic Ray Conference}
\CopyRight{\copyright 2003 by Universal Academy Press, Inc.}
%\tableofcontents
\pagenumbering{arabic}

\chapter{%   %%%%%%%%% <===== TITLE of the contribution
%%%%%%%%%%% The first letter of each word should be capital letter.
Asymmetries Observed In Giant Air Showers Using Water Cherenkov
Detectors.}

\author{%
%
% You can include as many co-authors as you wish, unless
% the title/author information fits within 1 page.
%
Mar\'{\i}a Teresa Dova,$^1$ for the Pierre Auger Collaboration $^2$ \\
{\it (1) Physics Department, Universidad Nacional de La Plata, C.C. 67 - 1900 La Plata, Argentina\\
(2) PAO, Av. San Martin Norte 304, (5613)
Malargue, Argentina} \\
}%% end of author

\section*{Abstract}

%The Auger Observatory will contain an array of 1600 particle detectors (SD),
%spread over 3000 km$^2$, overlooked by four fluorescence telescope
%stations. The surface array compromises water-Cherenkov detectors, on a
%1.5 km hexagonal grid.

Evidence of azimuthal asymmetries in the time
structure and signal size has been found in non-vertical showers as a
function of zenith angle. These asymmetries arise because of the different
paths traveled by particles in the upper and lower sides of the plane
perpendicular to the shower axis to reach detectors at the same axial
distances. The shower particles are differentially attenuated as they
traverse the atmosphere. Furthermore, most particles are not propagating
strictly in the shower direction but are on average going away from the
axis. This geometrical projection effect also contributes to the final
asymmetry.
These novel observations must be understood for parameterisation of the
lateral distribution function. Additionally, the asymmetry in time
distributions offers a new possibility for the determination of the mass
composition because its magnitude is strongly dependent on the fraction of
electromagnetic signal at the observation level. The asymmetries found in
data collected from the Engineering Array of the Auger Observatory will be
compared with Monte Carlo data.

\section{Introduction}

One of the foremost issues in astrophysics today is the origin of the ultra high energy cosmic rays (UHECR). An important ingredient to understand the origin of these energetic particles is the chemical composition of the UHECR detected. The Pierre Auger Observatory (PAO), currently under construction in Province of Mendoza, Argentina, is an international effort to make a high statistics study of the upper-end of the cosmic ray spectrum. An ``Engineering Array'' (EA), consisting of 32 water Cherenkov detectors  and two prototype fluorescence telescopes overlooking the ground detectors, has been operating in a steady state since January 2002. 

 In this paper we present a preliminary analysis of one aspect of the data collected by the 
ground array of the EA of the Auger Observatory at zenith angles smaller than 60$^0$. This is relevant for understanding azimuthal asymmetries observed in particle densities and arrival time distributions
studies, which are evident when one considers non-vertical showers [1]

\section{Asymmetries in EAS showers: old ideas, new approach.}

%The event reconstruction performed in all air shower arrays
%assume that the observation depends only on perpendicular distance
%from the shower axis.

The circular symmetry of the observed particle densities from vertical showers is
broken when considering inclined showers.  At a given distance from the shower core, the density of particles in the shower plane is higher on the side before the shower core impact point (``early region'') than on the opposite side (``late region''), mainly for photons and electrons.  There is a combination of effects which allow us to explain the asymmetries in the integrated signals of each detector in an event.

On the one hand particles hit detectors closer to the vertical in the ``early'' part of the shower
due to the fact that they diverge at larger angles from the shower axis, compared to the detectors in the ``late'' region. The usual projection of the ground densities into the shower plane results in an asymmetry (geometrical effect). At  moderate zenith angles ($\theta$ less than 40 degrees)
the asymmetry is dominated by this purely geometrical projection effect which
is largely compensated by the signal obtained from the particles hitting the  side area of the detectors [2].
On the other hand, there is an asymmetry arising from the evolution of the lateral distribution and the attenuation of the particles traveling longer paths in the atmosphere in the upper than the lower side of the plane
perpendicular to the shower axis (attenuation effects), largely discussed in [3]. The muon component in an EAS ($E_\mu \approx 1$ GeV at ground level) is typically attenuated and scattered less than the electromagnetic component ($E_{el} \approx 10$ MeV at ground level). This attenuation of the electromagnetic particles leads to the  observed asymmetries at large zenith angles. 
To illustrate: the path difference traveled by the particles in the ``late'' region at 1300 m from the shower core in the plane of the shower at  $\theta = 40^{\circ}$ is estimated to be 300 g cm$^{-2}$. Analysis of lateral distribution of $10^{19}$ eV proton showers at different atmospheric depths using Monte Carlo data (AIRES [4] with QGSJET01 [5] convoluted with the response of Auger ground detectors), gives a rough indication of the attenuation in the atmosphere. The electromagnetic component (EM) at ground level is attenuated by a factor of 2 with respect to an observing level at 300 g cm$^{-2}$ above ground. It is worth recalling at this point that the photon attenuation length in air for a 10 MeV photon, is estimated to be 45-50 g cm$^{-2}$ [6].

The muon component tends
to arrive earlier and over a shorter period of time than the electromagnetic one. To understand observed asymmetries in arrival distributions due to attenuation of electromagnetic particles in the ``late'' region, we have to take into account that, (i) the ratio of muon to EM components changes with primary composition and, (ii) it is dependent on distance from the core. We performed a preliminary analysis of the timing information of EAS using proton and iron simulation data to study the sensitivity of the Auger Observatory to these two components. We use typically three quantities: rise-time (time between 10\% and 50\% of the integrated signal), fall-time (time between 50\% and 90\%) and bulk-time (time between 10\% and 90\%). Functional fits to a linear cosine have been carried out to the timing variables as a function of azimuth angle in the shower plane ($\zeta$) at fixed range of core distances and the whole range for proton and iron shower at 35$^{\circ}$ and 60$^{\circ}$, as shown in Figure 1. The incoming direction of the shower is defined as $\zeta = 0^{\circ}$ ($-180^{\circ} \le \zeta \le 180^{\circ} $).
Shower fluctuations (resulting from fluctuations of the first interaction point), represented by the error bars in the plots,  are smaller for iron than proton showers as expected. Analysis using larger Monte Carlo statistics as well as a study of variations of the fitted parameters from the functional form $\tau = a + b \cos (\zeta)$  are underway. A first analysis seems to indicate that the fall-time would be a better discriminating tool.

\begin{figure}[t]
  \begin{center}
    \includegraphics[height=10.5pc]{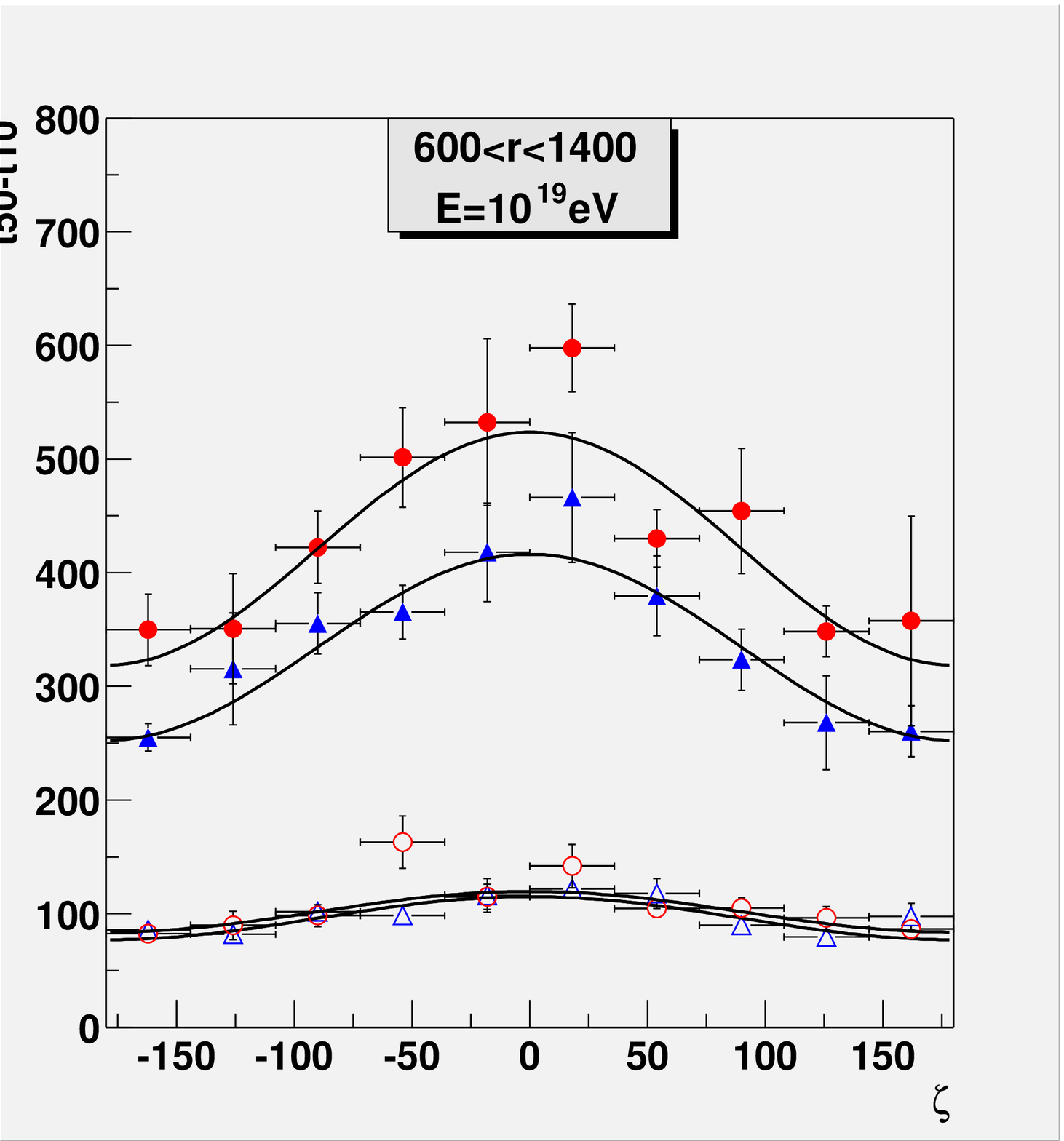} 
    \includegraphics[height=10.5pc]{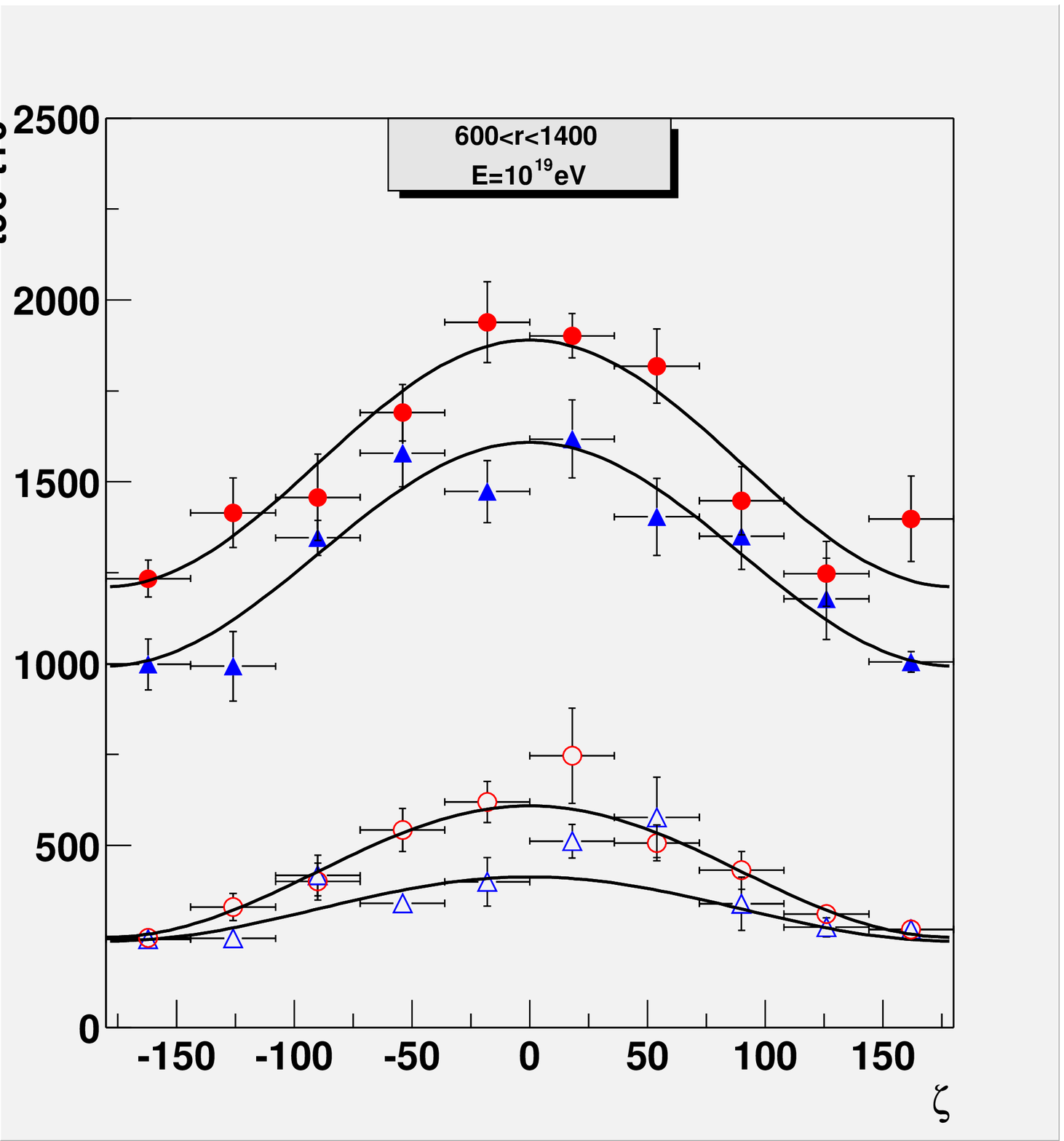} 
     \includegraphics[height=10.5pc]{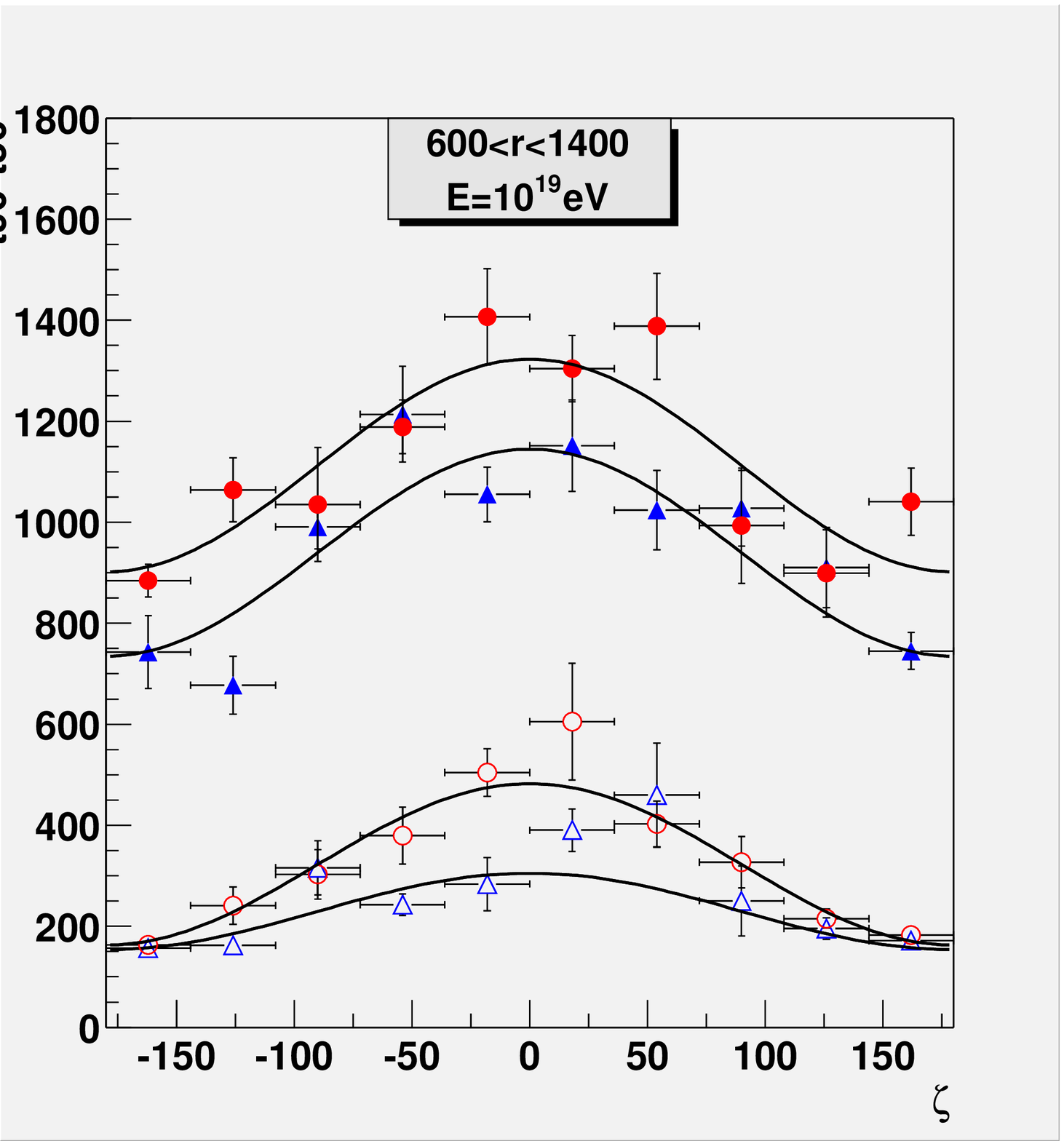} 
  \end{center}
  \vspace{-0.5pc}
  \caption{Rise-time (left), bulk-time (middle) and  fall-time (right) as a function of $\zeta$. Filled symbols: $35^\circ$, open symbols: $60^\circ$. Triangle: iron, Circle: proton. The error bars indicate the shower fluctuations. }
\end{figure}

\section{Preliminary analysis}

We expect stronger asymmetries at intermediate core distances, where the electromagnetic components dominates. A detailed analysis of a 15-fold event \#215819 has been performed, the zenith angle is $\approx60^\circ$ . The lateral distribution was determined fitting an NKG-like function with azimuthal dependence following the procedure in [3]. This kind of analysis can be only done with high multiplicity stations events. In Figure 2(top-left) we show the fitted lateral distribution. The corresponding distribution in azimuth angle, $\zeta$, normalized to the value of the density at $\zeta = \pi/2$, where the signal density preserves the symmetry, in the range 500 to 2000 m from the shower axis is displayed at the bottom-left. Dots with error bars, correspond to the data while the open dots are the result of the fit. The $\zeta$ dependence of the signal is evident. The plot on the top-right of Figure 2 shows the mean rise-time as a function of $\zeta$ for events with energy above 1 EeV collected in the period May to November 2002, in the radial range 600-1400 m. The corresponding azimuth dependence of the fall-time is presented in the last plot. The asymmetry is larger in this latest case as pointed out in section 2. 
\begin{figure}[t]
  \begin{center}
    \includegraphics[height=20pc,width=35pc]{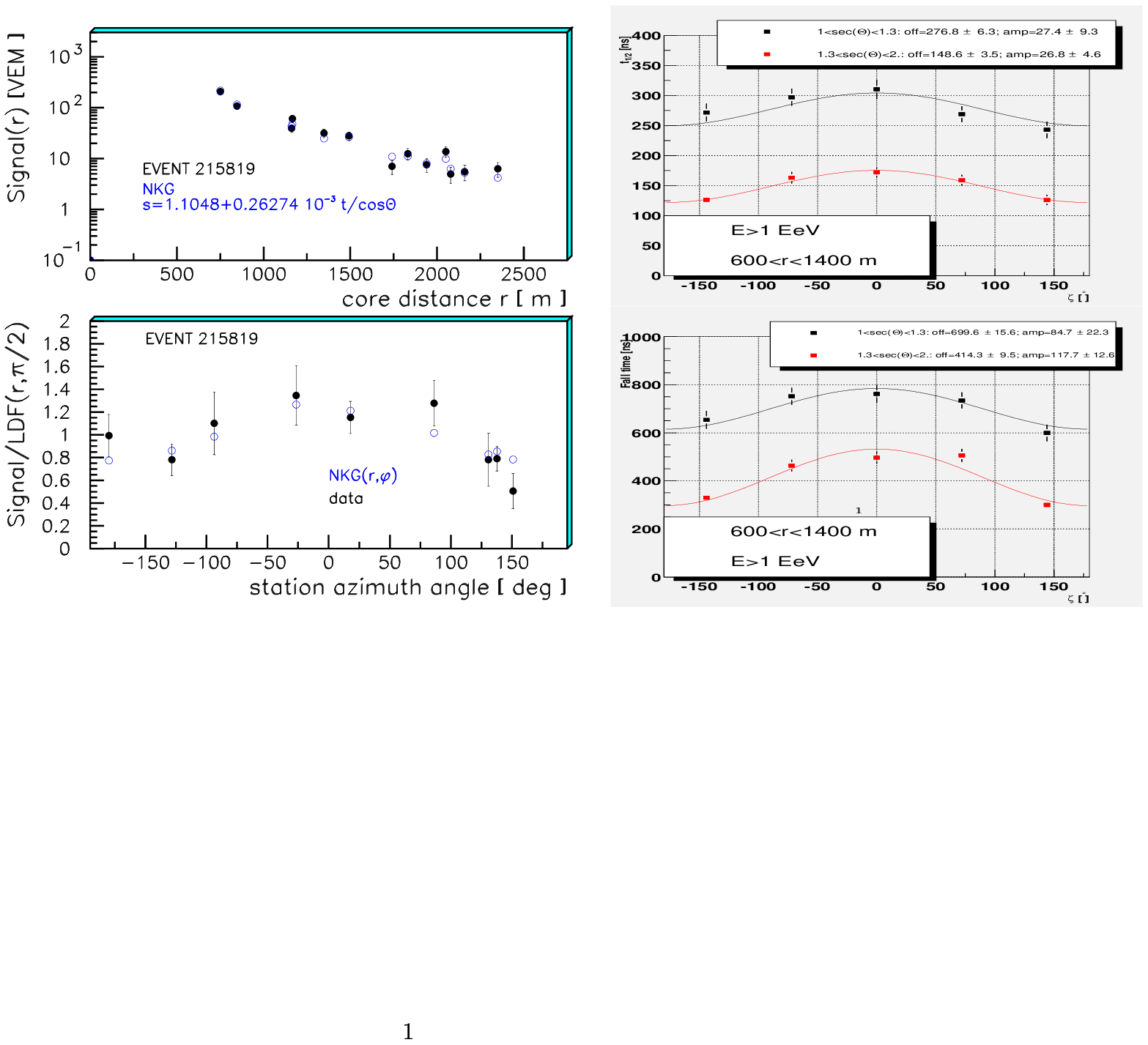}
  \end{center}
  \vspace{-0.5pc}
  \caption{Left: Integrated signal as a function of core distance and azimuth angle. Dots with error bars correspond to data. Open dots are the result of the fit. Right: Rise-time and fall-time as a function of $\zeta$.}
\end{figure}

All in all, the novel observations of asymmetries from the time 
distribution of the signals in the SD detectors promise rich 
information on the properties of EAS. This effect is important for determining the lateral distribution function and, additionally, 
it provides a new technique for primary mass separation.

\section{References}
\vspace{\baselineskip}
\re
1. Ave M. et al, Auger Note (GAP-2002-020), www.auger.org/admin
\re
2.\ Billoir P. et al, Auger Note (GAP-2000-017/2002-074), www.auger.org/admin
\re
3.\ Dova, M.T. et al, Astroparticle Physics 2003 18, 351
\re
4. Sciutto S.J., astroph/9911331
\re
5. Kalmykov N.N. et al, Yad.Fiz 1993, 56 105; Phys. Atom. Nucl. 1993, 56 346; Bull Russ. Acad.Sci (Physics) 1994, 58 1966.
\re
6.\ Hagiwara K. et al, Review of Particle Physics, 2002, Phys. Rev. D66, 010001

\endofpaper
\end{document}